\documentclass{elsart}
\usepackage{graphicx}
\begin{document}

\begin{frontmatter}
\begin{flushright}
\Large TTP03-10\\
\Large SFB/CPP-03-05\\
\Large THEP 03/05\\
\Large hep-ph/0303113\\
\end{flushright}
\title{Three-loop anomalous dimension\\
of the heavy-light quark current in HQET}
\author{K.G.~Chetyrkin}${}^{a,b,1}$\thanks{Permanent address:
Institute for Nuclear Research, Russian Academy of Sciences,
60th October Anniversary Prospect 7a, Moscow 117312, Russia.} 
\ead{chet@particle.uni-karlsruhe.de},
\author{A.G.~Grozin}${}^{b,2}$\thanks{Permanent address:
Budker Institute of Nuclear Physics, Novosibirsk, Russia.}
\ead{grozin@particle.uni-karlsruhe.de}
\address{${}^a$\it Albert-Ludwigs-Universit{\"a}t Freiburg,
D-79104 Freiburg, Germany }
\address{${}^b$Institut f\"ur Theoretische Teilchenphysik, Universit\"at Karlsruhe}
\begin{abstract}
The anomalous dimension of the heavy-light quark current in HQET
is calculated with three-loop accuracy,
as well as the renormalized heavy-quark propagator.
The NNL perturbative correction to $f_B/f_D$ is obtained.
\end{abstract}

\end{frontmatter}

\section{Introduction}
\label{Intro}

\begin{sloppypar}
Heavy quark physics progresses fast, both experimentally and theoretically.
Extraction of fundamental Standard Model parameters from high-precision data
from the $B$-factories requires thorough theoretical understanding
of strong-interaction effects.
In particular, higher radiative and power corrections
have to be calculated.
\end{sloppypar}

Understanding properties of hadrons with a heavy quark
has been boosted by the Heavy Quark Effective Theory
(HQET, see the textbook~\cite{MW:00}).
It is an effective field theory approximating QCD
for problems with a single heavy quark
when its 4-velocity $v$ remains approximately constant,
so that its momentum is $p=mv+k$,
and the characteristic residual momentum of the heavy quark $k$
is small compared to its mass $m$,
and characteristic momenta of light quarks and gluons are also small.
The HQET Lagrangian can be systematically constructed as a series in $1/m$.
Operators of full QCD are also expanded in $1/m$,
coefficients are HQET operators with the appropriate quantum numbers.

Here we are going to consider QCD heavy-light bilinear quark currents.
They have numerous applications:
vector and axial (with the anticommuting $\gamma_5$) currents
describe heavy-to-light transitions with $W^\pm$ emission,
tensor currents -- the $b\to s\gamma$ transition
which appears at one loop in the Standard Model.
The QCD heavy-light current is
$j_0=\bar{q}_0\Gamma Q_0=Z_j(\alpha_s^{(n_f)}(\mu))j(\mu)$,
where $\Gamma$ is a Dirac matrix,
and $n_f=n_l+1$ is the full number of flavours including the heavy one.
Its HQET expansion is
\begin{equation}
j(\mu') = C(\mu',\mu) \tilde{\jmath}(\mu)
+ \frac{1}{2m} \sum_i B^\Gamma_i(\mu',\mu) O_i(\mu)
+ \mathcal{O}(1/m^2)\,,
\label{Exp}
\end{equation}
where the HQET current is
$\tilde{\jmath}_0=\bar{q}_0\Gamma h_{v0}=\tilde{Z}_j(\alpha_s(\mu))\tilde{\jmath}(\mu)$,
$h_v=\rlap/v h_v$ is the HQET heavy-quark field,
and $O_i$ are dimension-4 HQET operators having the appropriate quantum numbers.
In order to have just one leading-order term in~(\ref{Exp}),
we should treat $\rlap/v$ and $\gamma_\bot^\mu=\gamma^\mu-\rlap/v v^\mu$
in $\Gamma$ separately.

The dependence of the matching coefficient $C(\mu',\mu)$
on the normalization scales $\mu'$ and $\mu$
is determined by the renormalization group:
\begin{equation}
C(\mu',\mu) = C(m,m) \exp \Biggl[
\int\limits_{\alpha_s^{(n_f)}(m)}^{\alpha_s^{(n_f)}(\mu')}
\frac{\gamma_j(\alpha_s)}{2\beta^{(n_f)}(\alpha_s)}
\frac{d\alpha_s}{\alpha_s}
- \int\limits_{\alpha_s^{(n_l)}(m)}^{\alpha_s^{(n_l)}(\mu)}
\frac{\tilde{\gamma}_j(\alpha_s)}{2\beta^{(n_l)}(\alpha_s)}
\frac{d\alpha_s}{\alpha_s}
\Biggr]\,,
\label{RG}
\end{equation}
where
\begin{eqnarray*}
&&\tilde{\gamma}_j = \frac{d\log\tilde{Z}}{d\log\mu}
= \tilde{\gamma}_{j0} \frac{\alpha_s}{4\pi}
+ \tilde{\gamma}_{j1} \left(\frac{\alpha_s}{4\pi}\right)^2
+ \cdots\\
&&\beta^{(n_l)} = - \frac{1}{2} \frac{d\log\alpha_s^{(n_l)}}{d\log\mu}
= \beta_0^{(n_l)} \frac{\alpha_s}{4\pi}
+ \beta_1^{(n_l)} \left(\frac{\alpha_s}{4\pi}\right)^2
+ \cdots
\end{eqnarray*}
and $\alpha_s^{(n_l)}(\mu)$ contains only $n_l$ light flavours.
The HQET current anomalous dimension $\tilde{\gamma}_j$
does not depend on the Dirac structure of the current $\Gamma$,
and was known at one~\cite{VS:87,PW:88} and two~\cite{JM:91,BG:91} loops.
The QCD current anomalous dimension $\gamma_j$
vanishes for the vector and axial currents
(with the anticommuting $\gamma_5$) to all orders.
The QCD anomalous dimensions of the scalar and pseudoscalar 
(with the anticommuting $\gamma_5$) currents  are known to be
equal (modulo a sign) to the quark mass anomalous dimension,
the latter is available at four loops~\cite{Chetyrkin:1997dh,Vermaseren:1997fq}.
The anomalous dimensions of the axial and pseudoscalar currents
with the 't~Hooft--Veltman $\gamma_5$ are known at three loops~\cite{La:93}.
Generic expressions for an arbitrary $\Gamma$
are known at two~\cite{BG:95} and three~\cite{Gr:00} loops.
The matching coefficients
$C(m,m)=1+c_1\alpha_s^{(n_l)}(m)/(4\pi)+c_2[\alpha_s^{(n_l)}(m)/(4\pi)]^2+\cdots$
contain no large logarithms;
for all $\Gamma$, they are known at one~\cite{EH:90}
and two~\cite{BG:95,G:98} loops.

As one can see from~(\ref{RG}),
two-loop (next-to-next-to-leading order, NNL) corrections to $C(m,m)$ should be used
together with three-loop (NNL) anomalous dimensions and $\beta$-function.
Therefore, the results of~\cite{BG:95,G:98} cannot be fully utilized
without knowing $\tilde{\gamma}_{j2}$.%
\footnote{$\tilde{\gamma}_j$ cancels in ratios of matrix elements,
which were investigated in~\cite{BG:95}.}
The aim of this work is to calculate it.

The $B$-meson leptonic decay constant $f_B$ is one of the most important
non-perturbative parameters in $B$ physics.
It has not been measured experimentally;
observation of the $B^-\to\tau^-\bar{\nu}_\tau$ decay
is a very difficult (though not hopeless) task for $B$-factories.
Theoretical estimates vary widely.
Improving their accuracy is extremely important.
The HQET matrix element ${<}0|\tilde{\jmath}(\mu)|B{>}$
at a low $\mu\ll m_b$ can be obtained by some non-perturbative technique,
like lattice simulations (see recent review talks~\cite{lat})
or HQET sum rules~\cite{BG:92,BBBD:92}.
Direct lattice calculations with a heavy quark of mass $\sim m_b$
will not be possible in the foreseeble future,
because they require the lattice spacing $a\ll1/m_b$.
A way to avoid the far extrapolation from $m\sim m_c$
(which can be simulated on present-day lattices) to $m\sim m_b$
is to use the lattice HQET, in which the heavy-quark mass $m$
does not appear.
Such simulations produce results normalized at a low scale $\mu\sim1/a$.
They should be matched to the continuum HQET
and run up to $\mu=m_b$ using the three-loop
anomalous dimension $\tilde{\gamma}_j$, and then matched to QCD
using the two-loop $C(m,m)$~\cite{BG:95,G:98}.
Therefore, calculation of $\tilde{\gamma}_{j2}$ is one of the ingredients
necessary to make lattice evaluations of $f_B$ more precise.
Similarly, HQET sum rules produce results normalized at low $\mu$
of the order of a typical Borel parameter,
they also have to be run up to $\mu=m_b$.
The knowledge of $\tilde{\gamma}_j$ is also important
for improving the accuracy of lattice and sum-rule calculations
of the form factors of heavy-to-light exclusive semileptonic decays
$B\to\pi$, $B\to\rho$ (for soft $\pi$, $\rho$)
and exclusive rare decays caused by the $b\to s\gamma$ transition.

\section{Three-loop HQET propagator diagrams}
\label{Dia}

A systematic method of calculation of two-loop HQET propagator diagrams
based on integration by parts~\cite{CT:81} has been constructed in~\cite{BG:91}.
At three loops, this has been done in~\cite{G:00}.
The algorithm has been implemented as a REDUCE package Grinder available at
http://www-ttp.physik.uni-karlsruhe.de/\\Progdata/ttp00/ttp00-01/.

\begin{figure}[ht]
\begin{center}
\begin{picture}(132,114)
\put(66,59){\makebox(0,0){\includegraphics{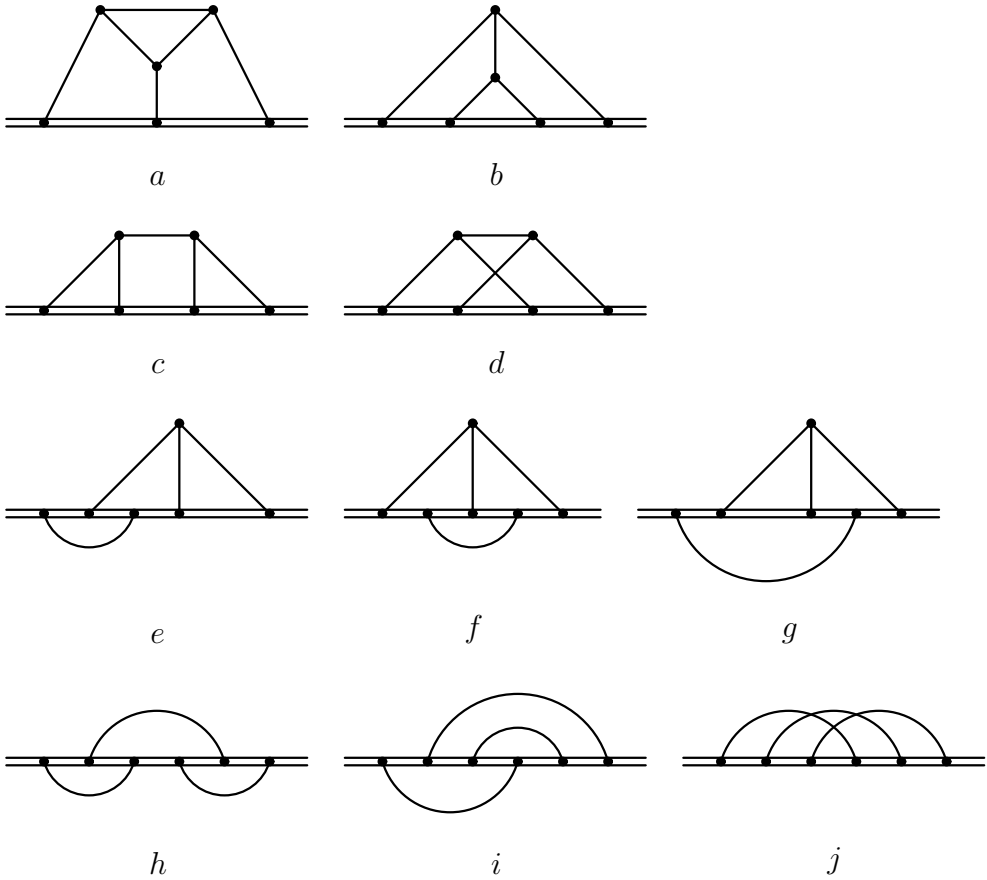}}}
\put(21,89){\makebox(0,0)[b]{$a$}}
\put(66,89){\makebox(0,0)[b]{$b$}}
\put(21,64){\makebox(0,0)[b]{$c$}}
\put(66,64){\makebox(0,0)[b]{$d$}}
\put(21,28){\makebox(0,0)[b]{$e$}}
\put(63,28){\makebox(0,0)[b]{$f$}}
\put(105,28){\makebox(0,0)[b]{$g$}}
\put(21,-2.75){\makebox(0,0)[b]{$h$}}
\put(66,-2.75){\makebox(0,0)[b]{$i$}}
\put(111,-2.75){\makebox(0,0)[b]{$j$}}
\end{picture}
\end{center}
\caption{Topologies of three-loop propagator diagrams in HQET}
\label{Top}
\end{figure}

\begin{figure}[ht]
\begin{center}
\begin{picture}(122,67.5)
\put(61,33.75){\makebox(0,0){\includegraphics{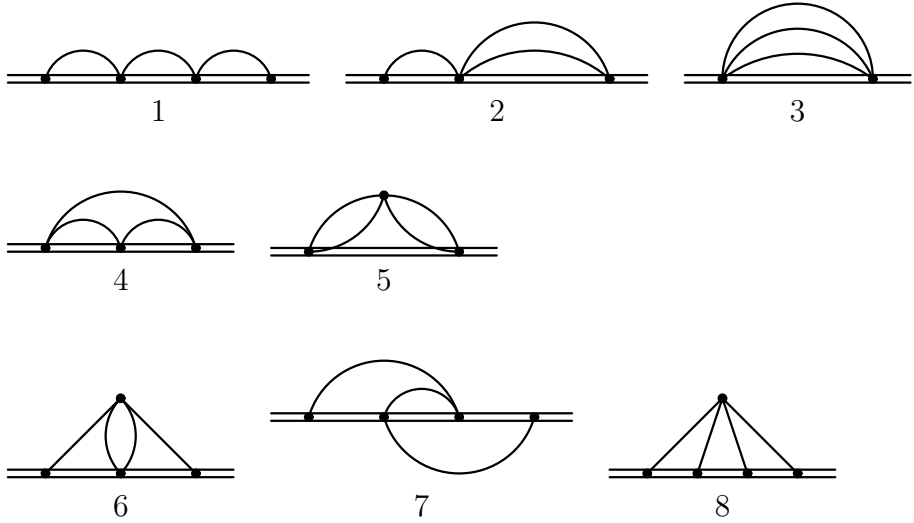}}}
\put(21,49.5){\makebox(0,0)[b]{1}}
\put(66,49.5){\makebox(0,0)[b]{2}}
\put(106,49.5){\makebox(0,0)[b]{3}}
\put(16,27){\makebox(0,0)[b]{4}}
\put(51,27){\makebox(0,0)[b]{5}}
\put(16,-3){\makebox(0,0)[b]{6}}
\put(56,-3){\makebox(0,0)[b]{7}}
\put(96,-3){\makebox(0,0)[b]{8}}
\end{picture}
\end{center}
\caption{Basis integrals}
\label{Basis}
\end{figure}

There are 10 generic topologies of three-loop propagator diagrams in HQET
(Fig.~\ref{Top}).
Grinder reduces all of them to linear combinations of 8 basis integrals
shown in Fig.~\ref{Basis}, with coefficients being rational functions
of the space-time dimension $d$.

The first 5 basis integrals are trivial:
\begin{equation}
B_1 = I_1^3\,,\quad
B_2 = I_1 I_2\,,\quad
B_3 = I_3\,,\quad
B_4 = I_3 \frac{I_1^2}{I_2}\,,\quad
B_5 = I_3 \frac{G_1^2}{G_2}\,,
\label{B15}
\end{equation}
where
\begin{eqnarray*}
&&I_n = \frac{\Gamma(1+2n\varepsilon)\Gamma^n(1-\varepsilon)}{(1-n(d-2))_{2n}}\,,\\
&&G_n =
\frac{1}{\left(n+1-n\frac{d}{2}\right)_n \left((n+1)\frac{d}{2}-2n-1\right)_n}\,
\frac{\Gamma(1+n\varepsilon)\Gamma^{n+1}(1-\varepsilon)}{\Gamma(1-(n+1)\varepsilon)}
\end{eqnarray*}
are the $n$-loop sunset HQET and massless integrals.

\begin{figure}[ht]
\begin{center}
\begin{picture}(102,26)
\put(51,13){\makebox(0,0){\includegraphics{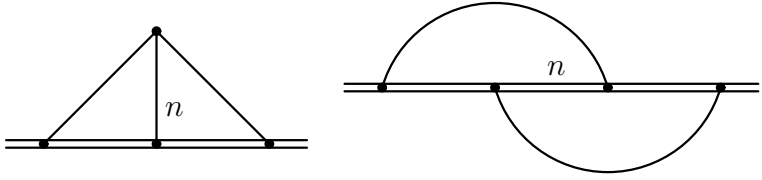}}}
\put(22,10){\makebox(0,0)[l]{$n$}}
\put(74,14.5){\makebox(0,0)[b]{$n$}}
\end{picture}
\end{center}
\caption{Two-loop integrals $I$ and $J$}
\label{F67}
\end{figure}

The next two basis integrals are
\begin{equation}
B_6 = G_1 I(1,1,1,1,\varepsilon)\,,\quad
B_7 = I_1 J(1,1,-1+2\varepsilon,1,1)\,.
\label{B67}
\end{equation}
We use recurrence relations for the two-loop integrals $I$ and $J$
(see~\cite{G:00})
to re-express them via $I(1,1,1,1,1+\varepsilon)$ and $J(1,1,2+2\varepsilon,1,1)$.
The integral $I(1,1,1,1,n)$ for any $n$ was calculated in~\cite{BB:94},
and $J(1,1,n,n_1,n_2)$ for any $n$, $n_1$, $n_2$ -- in~\cite{G:00} (Fig.~\ref{F67}).
In particular,
\begin{eqnarray}
&&I(1,1,1,1,1+\varepsilon) =
\frac{4\Gamma(1-\varepsilon)}{9(d-3)(d-4)^2}
\Biggl[ \frac{\Gamma(1+4\varepsilon)\Gamma(1+6\varepsilon)}{\Gamma(1+7\varepsilon)}
\nonumber\\
&&\quad{}\times
{}_3\!F_2 \Biggl(
\begin{array}{c}
3\varepsilon, 3\varepsilon, 6\varepsilon\\ 1+3\varepsilon, 1+7\varepsilon
\end{array}
\Biggr.\Biggl|\, 1 \Biggr)
- \Gamma^2(1+3\varepsilon) \Gamma(1-3\varepsilon)
\Biggr]\,,
\nonumber\\
&&J(1,1,2+2\varepsilon,1,1) =
\frac{1}{3(d-4)(d-5)(d-6)(2d-9)}
\frac{\Gamma(1+6\varepsilon)\Gamma^2(1-\varepsilon)}{\Gamma(1+2\varepsilon)}
\nonumber\\
&&{}\times{}_3\!F_2 \Biggl(
\begin{array}{c}
1, 2-2\varepsilon, 1+4\varepsilon\\ 3+2\varepsilon, 2+4\varepsilon
\end{array}
\Biggr.\Biggl|\, 1 \Biggr)\,.
\label{IJ}
\end{eqnarray}
Such hypergeometric functions can be easily expanded in $\varepsilon$
in terms of multiple $\zeta$ sums using the C++ library nestedsums~\cite{We:02}.
Reducing them to the minimal basis~\cite{BBB:97}, we obtain
\begin{eqnarray}
&&{}_3\!F_2 \Biggl(
\begin{array}{c}
3\varepsilon, 3\varepsilon, 6\varepsilon\\ 1+3\varepsilon, 1+7\varepsilon
\end{array}
\Biggr.\Biggl|\, 1 \Biggr) =
1 + 54 \zeta_3 \varepsilon^3 - 513 \zeta_4 \varepsilon^4
+ 54 (25\zeta_5+28\zeta_2\zeta_3) \varepsilon^5
\nonumber\\
&&\quad{} + \cdots
\nonumber\\
&&{}_3\!F_2 \Biggl(
\begin{array}{c}
1, 2-2\varepsilon, 1+4\varepsilon\\ 3+2\varepsilon, 2+4\varepsilon
\end{array}
\Biggr.\Biggl|\, 1 \Biggr) =
2 + 6(-2\zeta_2+3) \varepsilon + 12(10\zeta_3-11\zeta_2+6) \varepsilon^2
\nonumber\\
&&\quad{} + 24(-28\zeta_4+55\zeta_3-27\zeta_2+9) \varepsilon^3
\nonumber\\
&&\quad{}
+ 48(94\zeta_5-16\zeta_2\zeta_3-154\zeta_4+135\zeta_3-45\zeta_2+12) \varepsilon^4 + \cdots
\label{Fexp}
\end{eqnarray}
Several more terms of these expansions can be found easily.

Finally, the most difficult basis integral $B_8$ can be found as following.
We calculate the ladder diagram $I_c(1,1,1,1,1,1,1,1)$
(Fig.~\ref{Top}$c$, see~\cite{G:00}) using Grinder, and solve for $B_8$.
This diagram is convergent; at $d=4$ it is related~\cite{CM:02}
by inversion of Euclidean integration momenta to a known~\cite{MR:00} on-shell diagram:
\begin{equation}
I_c(1,1,1,1,1,1,1,1) = - 5 \zeta_5 + 12 \zeta_2 \zeta_3
+ \mathcal{O}(\varepsilon)\,.
\label{F8}
\end{equation}

A REDUCE program which expands in $\varepsilon$ results produced by Grinder
can be obtained at~\cite{Progdata}.

\section{Heavy-quark propagator}
\label{Prop}

Due to the non-abelian exponentiation theorem~\cite{Ga:83,FT:84},
the bare HQET heavy-quark propagator in the coordinate space can be written as
\begin{eqnarray}
&&\tilde{S}(t) = -i\theta(t) \exp \Biggl[
C_F \frac{g_0^2}{(4\pi)^{d/2}} \left(\frac{it}{2}\right)^{2\varepsilon} S_F
+ C_F \frac{g_0^4}{(4\pi)^d} \left(\frac{it}{2}\right)^{4\varepsilon}
\left(C_A S_{FA} + T_F n_l S_{Fl}\right)
\nonumber\\
&&\quad{}
+ C_F \frac{g_0^6}{(4\pi)^{3d/2}} \left(\frac{it}{2}\right)^{6\varepsilon}
\left(C_A^2 S_{FAA} + C_F T_F n_l S_{FFl} + C_A T_F n_l S_{FAl}
+ \left(T_F n_l\right)^2 S_{Fll}\right)
\nonumber\\
&&\quad{} + \cdots \Biggr]\,.
\label{S0}
\end{eqnarray}
Let's see how this happens explicitely, up to three loops.

\begin{figure}[ht]
\begin{center}
\begin{picture}(107,38)
\put(53.5,19.5){\makebox(0,0){\includegraphics{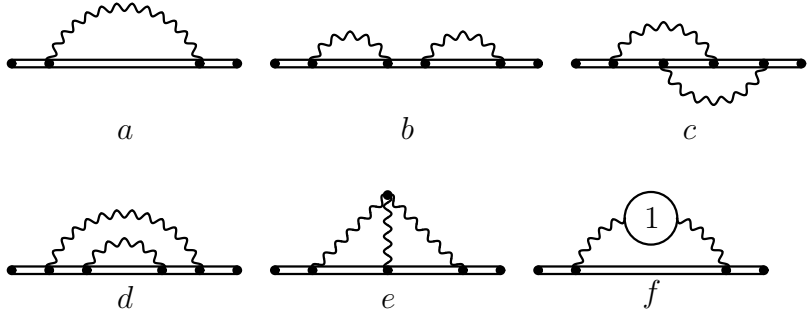}}}
\put(16,19.5){\makebox(0,0)[b]{$a$}}
\put(53.5,19.5){\makebox(0,0)[b]{$b$}}
\put(91,19.5){\makebox(0,0)[b]{$c$}}
\put(16,-3){\makebox(0,0)[b]{$d$}}
\put(51,-3){\makebox(0,0)[b]{$e$}}
\put(86,-3){\makebox(0,0)[b]{$f$}}
\put(86,9){\makebox(0,0){1}}
\end{picture}
\end{center}
\caption{One- and two-loop diagrams for the HQET propagator}
\label{Prop12}
\end{figure}

\begin{figure}[ht]
\begin{center}
\begin{picture}(142.2,54.9)
\put(71.1,27.45){\makebox(0,0){\includegraphics[scale=0.9]{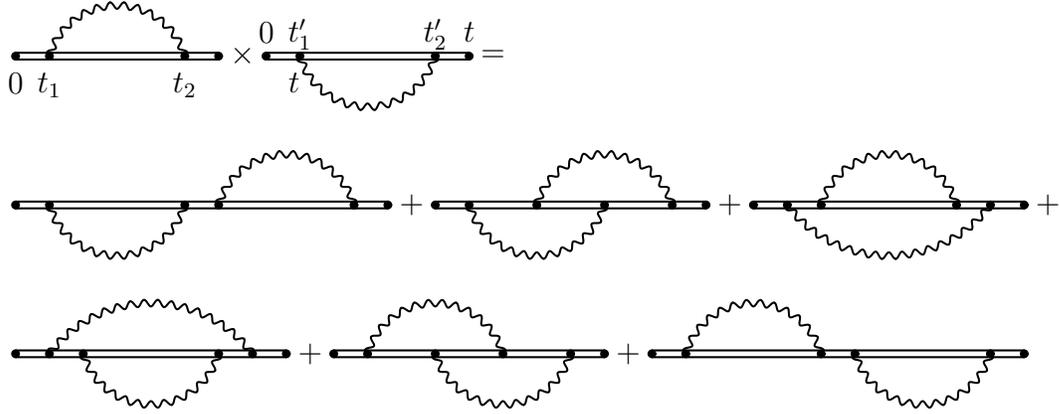}}}
\put(0.9,45){\makebox(0,0)[t]{0}}
\put(5.4,45){\makebox(0,0)[t]{$t_1$}}
\put(23.4,45){\makebox(0,0)[t]{$t_2$}}
\put(37.9,45){\makebox(0,0)[t]{$t$}}
\put(34.2,51.75){\makebox(0,0)[t]{0}}
\put(38.7,52.2){\makebox(0,0)[t]{$t_1'$}}
\put(56.7,52.2){\makebox(0,0)[t]{$t_2'$}}
\put(61.2,51.75){\makebox(0,0)[t]{$t$}}
\put(31.05,47.25){\makebox(0,0){$\times$}}
\put(64.35,47.25){\makebox(0,0){=}}
\put(53.55,27.45){\makebox(0,0){+}}
\put(95.85,27.45){\makebox(0,0){+}}
\put(138.15,27.45){\makebox(0,0){+}}
\put(40.05,7.65){\makebox(0,0){+}}
\put(82.35,7.65){\makebox(0,0){+}}
\end{picture}
\end{center}
\caption{Exponentiation theorem}
\label{Expo}
\end{figure}

Suppose we have calculated the one-loop correction to the HQET propagator
in the coordinate space (Fig.~\ref{Prop12}$a$).
Let's multiply this correction by itself (Fig.~\ref{Expo}).
We get an integral in $t_1$, $t_2$, $t_1'$, $t_2'$
with $0<t_1<t_2<t$, $0<t_1'<t_2'<t$.
Ordering of primed and non-primed integration times can be arbitrary.
The integration area is subdivided into 6 regions,
corresponding to 6 diagrams in Fig.~\ref{Expo}.
If the colour factors of the diagrams Fig.~\ref{Prop12}$c$, $d$
were the same as that of the one-particle-reducible diagram (Fig.~\ref{Prop12}$b$),
i.~e.\ equal to the square of the colour factor $C_F$
of the one-loop diagram (Fig.~\ref{Prop12}$a$) (as in the abelian case),
then the sum of the diagrams Fig.~\ref{Prop12}$b$, $c$, $d$ would be equal
to $\frac{1}{2}$ of the square of the one-loop correction $S_F$ (Fig.~\ref{Prop12}$a$),
as given by the square of the first term in the expansion of the exponent~(\ref{S0}).
In the non-abelian case, however, the colour factor of Fig.~\ref{Prop12}$c$
differs from $C_F^2$ by $-C_F C_A/2$,
which is the colour factor of Fig.~\ref{Prop12}$e$.
This is because when we reduce the colour factor of Fig.~\ref{Prop12}$c$
to that of the reducible diagram Fig.~\ref{Prop12}$b$,
we get an extra term with the commutator $[t^a,t^b]$,
which has the colour structure of Fig.~\ref{Prop12}$e$.
Therefore, we should include the contribution of Fig.~\ref{Prop12}$c$
with $-C_F C_A/2$ instead of its full colour factor into the term $S_{FA}$
in the exponent~(\ref{S0}).
This part of the colour factor is called maximally non-abelian~\cite{Ga:83}
or colour-connected~\cite{FT:84}.
Of course, the diagram Fig.~\ref{Prop12}$e$ also contributes to $S_{FA}$;
the diagram Fig.~\ref{Prop12}$f$, with the one-loop gluon self-energy correction,
contributes to $S_{Fl}$ (quark loop) and $S_{FA}$ (gluon and ghost loops).

\begin{figure}[ht]
\begin{center}
\begin{picture}(141.3,101.7)
\put(70.65,51.3){\makebox(0,0){\includegraphics[scale=0.9]{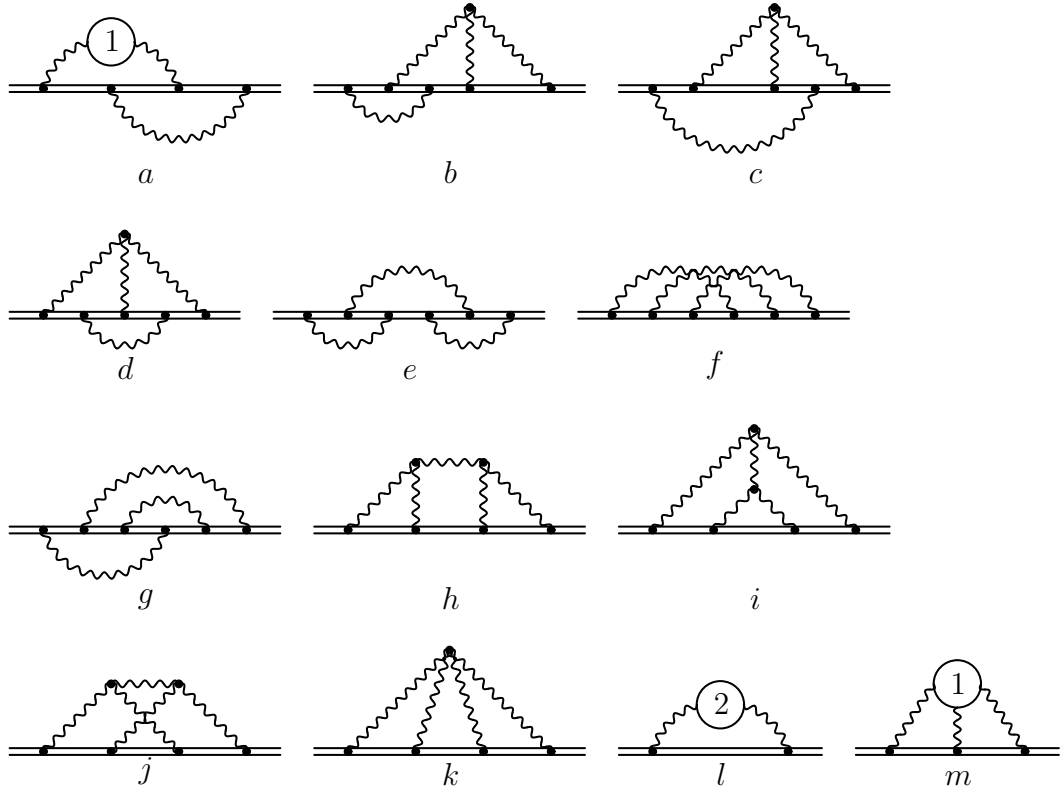}}}
\put(18.9,77.4){\makebox(0,0)[b]{$a$}}
\put(59.4,77.4){\makebox(0,0)[b]{$b$}}
\put(99.9,77.4){\makebox(0,0)[b]{$c$}}
\put(16.2,51.3){\makebox(0,0)[b]{$d$}}
\put(54,51.3){\makebox(0,0)[b]{$e$}}
\put(94.5,51.3){\makebox(0,0)[b]{$f$}}
\put(18.9,20.7){\makebox(0,0)[b]{$g$}}
\put(59.4,20.7){\makebox(0,0)[b]{$h$}}
\put(99.9,20.7){\makebox(0,0)[b]{$i$}}
\put(18.9,-2.7){\makebox(0,0)[b]{$j$}}
\put(59.4,-2.7){\makebox(0,0)[b]{$k$}}
\put(95.4,-2.7){\makebox(0,0)[b]{$l$}}
\put(126.9,-2.7){\makebox(0,0)[b]{$m$}}
\put(14.4,96.3){\makebox(0,0){1}}
\put(95.4,8.1){\makebox(0,0){2}}
\put(126.9,10.8){\makebox(0,0){1}}
\end{picture}
\end{center}
\caption{Three-loop diagrams contributing to the three-loop term
in the exponent~(\ref{S0})}
\label{Prop3}
\end{figure}

In the same way, we can consider the set of three-loop diagrams
obtained by multiplying the corrections of Fig.~\ref{Prop12}$a$ and $f$.
We can imagine that this set is obtained from the one-particle-reducible diagram
by allowing the gluon -- heavy-quark vertices to slide along the heavy-quark line,
crossing each other.
These diagrams are said to contain two connected webs (or c-webs)~\cite{FT:84}.
Everything is already accounted for by the product
of the one-loop correction and the (Fig.~\ref{Prop12}$f$ part of) two-loop correction
in the expansion of the exponent~(\ref{S0}),
except the contribution of Fig.~\ref{Prop3}$a$ (and its mirror-symmetric),
taken with the maximally non-abelian part of its colour factor.
It contributes to the three-loop correction in the exponent.
Similarly, out of all the diagrams with two connected webs
Fig.~\ref{Prop12}$a$ and $e$,
only those of Fig.~\ref{Prop3}$b$, $c$ (plus their mirror-symmetric ones), and $d$
contribute to $S_{FAA}$, with the maximally non-abelian part of their colour factors.
This part appears, in the case of Fig.~\ref{Prop3}$b$, for example,
when we commute $t^a$ matrices to obtain the colour structure of the reducible diagram;
it is identical to the colour factor of Fig.~\ref{Prop3}$h$,
equal to $C_F C_A^2/4$.

Now we are going to consider all the diagrams with three gluons,
both ends of which are attached to the heavy-quark line.
They are said to contain three c-webs,
each of them is that of Fig.~\ref{Prop12}$a$.
We decompose the colour factors of these 15 diagrams in the following way.
We move the vertices along the heavy-quark line in such a way
as to disentangle those c-webs.
While doing so, we get extra terms from the commutators,
having colour structures of the corresponding diagrams with the three-gluon vertex.
These diagrams have fewer c-webs, which are more complicated.
Finally, each colour factor can be expressed as a linear combination of three ones:
$C_F^3$ (3 c-webs of Fig.~\ref{Prop12}$a$);
$-C_F^2 C_A/2$ (2 c-webs of Fig.~\ref{Prop12}$a$ and $e$);
$C_F C_A^2/4$ (1 c-web of Fig.~\ref{Prop3}$h$).
The first one occurs with the unit coefficient in all 15 colour factors.
The sum of the corresponding contributions is just the term with the cube
of the one-loop correction in the expansion of the exponent~(\ref{S0}).
The second colour structure occurs in the diagrams
obtained by multiplying Fig.~\ref{Prop12}$a$ and $c$;
the sum of the corresponding contributions is contained in the product
of the one-loop term and the two-loop one in the expansion of the exponent.
We are left with the colour-connected parts of the colour factors
(a single c-web contributions).
They are present in the diagrams Fig.~\ref{Prop3}$e$, $f$, $g$
(and its mirror-symmetric).
They contribute to $S_{FAA}$ in~(\ref{S0}).

\begin{figure}[ht]
\begin{center}
\begin{picture}(92,22)
\put(46,11){\makebox(0,0){\includegraphics{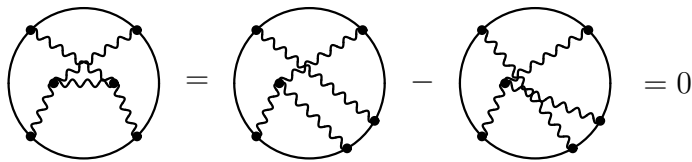}}}
\put(26,11){\makebox(0,0){${}={}$}}
\put(56,11){\makebox(0,0){${}-{}$}}
\put(88,11){\makebox(0,0){${}=0$}}
\end{picture}
\end{center}
\caption{Vanishing colour factor}
\label{Col0}
\end{figure}

We are left with the diagrams containing only a single connected web.
Those of Fig.~\ref{Prop3}$h$ and $i$ have equal colour factors
(this is evident if we close the quark line),
they contribute to $S_{FAA}$.
The colour factor of Fig.~\ref{Prop3}$j$ is zero.
This becomes clear if we close the quark line
and write a three-gluon vertex as the commutator (Fig.~\ref{Col0}).
The diagram with the four-gluon vertex (Fig.~\ref{Prop3}$k$)
can be decomposed into three terms,
with colour factors of Fig.~\ref{Prop3}$h$, $i$, $j$.
The diagram Fig.~\ref{Prop3}$l$ contains two-loop gluon self-energy corrections,
including one-particle-reducible ones;
it contributes to $S_{FAA}$, $S_{FFl}$, $S_{FAl}$, $S_{Fll}$.
The diagram Fig.~\ref{Prop3}$m$ contains one-loop corrections
to the three-gluon vertex, including one-particle-reducible ones
(i.~e., one-loop self-energy corrections to each gluon propagator
in Fig.~\ref{Prop12}$e$);
it contributes to $S_{FAA}$, $S_{FAl}$.

In order to obtain the heavy-quark field renormalization constant $\tilde{Z}_Q$,
we should re-express this bare propagator
in terms of the $\overline{\mathrm{MS}}$ renormalized coupling $\alpha_s(\mu)$ and
gauge-fixing parameter $a(\mu)$ instead of $g_0^2$ and $a_0$:%
\footnote{$a_0$ is defined so that the bare gluon propagator is
$\left[g_{\mu\nu}-(1-a_0)p_\mu p_\nu/p^2\right]/p^2$.}
\[
\frac{g_0^2}{(4\pi)^{d/2}} = \mu^{2\varepsilon} \mathrm{e}^{\gamma_E\varepsilon}
\frac{\alpha_s(\mu)}{4\pi} Z_\alpha(\alpha_s(\mu))\,,\quad
a_0 = Z_A(\alpha_s(\mu)) a(\mu)\,,
\]
and require that the renormalized propagator
$\tilde{S}_{\mathrm{r}}(t;\mu)=\tilde{Z}_Q^{-1}\tilde{S}(t)$
is finite in the limit $\varepsilon\to0$.
After this re-expression, $S(t)$ still has the exponential form
with the same colour structures as in~(\ref{S0}).
Therefore, the renormalization constant can be written in the exponential form, too:
\begin{eqnarray}
&&\tilde{Z}_Q = \exp \Biggl[ C_F \frac{\alpha_s}{4\pi} Z_F
+ C_F \left(\frac{\alpha_s}{4\pi}\right)^2 \left(C_A Z_{FA} + T_F n_l Z_{Fl}\right)
\label{ZQ}\\
&&{} + C_F \left(\frac{\alpha_s}{4\pi}\right)^3
\left(C_A^2 Z_{FAA} + C_F T_F n_l Z_{FFl} + C_A T_F n_l Z_{FAl}
+ \left(T_F n_l\right)^2 Z_{Fll}\right)
+ \cdots \Biggr]\,.
\nonumber
\end{eqnarray}
Hence the heavy-quark field anomalous dimension
$\tilde{\gamma}_Q=d\log\tilde{Z}_Q/d\log\mu$
has no $C_F^2$ term at two loops
and no $C_F^3$, $C_F^2 C_A$ terms at three loops.

We have calculated the HQET heavy-quark self-energy
$\tilde{\Sigma}(\omega)$ with the three-loop accuracy, in an arbitrary
covariant gauge.  We have used GEFICOM~\cite{geficom} for automatic
preparation of input data for Grinder. GEFICOM used QGRAF~\cite{No:93}
for generation of diagrams.  Then a specially written Mathematica
program classified them into topologies, established the momentum
routing required by Grinder for each topology.  Dirac traces, index
contractions and colour factors were performed with  FORM~\cite{Ve:00}.
Then Feynman integrals were calculated with the REDUCE package Grinder~\cite{G:00}.
The whole process is completely automatic.
We have checked that the bare propagator obtained from this self-energy,
after Fourier transforming to the coordinate space,
obeys exponentiation~(\ref{S0}), exactly at any $a_0$ and $d$.
The complete result for $\tilde{S}(\omega)$,
as well as its expansion in $\varepsilon$,
can be obtained at~\cite{Progdata}.

Finally, we obtain the anomalous dimension
\begin{eqnarray}
&&\tilde{\gamma}_Q = 2 C_F (a-3) \frac{\alpha_s}{4\pi}
+ C_F \left[C_A \left(\frac{a^2}{2}+4a-\frac{179}{6}\right) + \frac{32}{3} T_F n_l \right]
\left(\frac{\alpha_s}{4\pi}\right)^2
\nonumber\\
&&\quad{} + C_F \Biggl[
C_A^2 \biggl(\frac{5}{8}a^3 + \left(\frac{3}{4}\zeta_3+\frac{39}{16}\right)a^2
+ \left(-8\zeta_4+6\zeta_3+\frac{271}{16}\right)a
\nonumber\\
&&\qquad{} - 24\zeta_4 - \frac{123}{4}\zeta_3 - \frac{23815}{216} \biggr)
\nonumber\\
&&\quad{} + 6 C_F T_F n_l (-16\zeta_3+17)
+ C_A T_F n_l \left(-\frac{17}{2}a+96\zeta_3+\frac{782}{27}\right)
\nonumber\\
&&\quad{} + \frac{160}{27} \left(T_F n_l\right)^2
\Biggr] \left(\frac{\alpha_s}{4\pi}\right)^3
+ \cdots
\label{gammaQ}
\end{eqnarray}
It has the structure required by exponentiation.
In the abelian theory without light fermions (Bloch--Nordsieck model),
it equals the one-loop term exactly;
it vanishes in the Yennie gauge $a=3$
(which has been introduced for exactly this reason).
The two-loop term coincides with~\cite{BG:91}.
The three-loop term with $(T_F n_l)^2$
is contained in the all-order large-$n_l$ result~\cite{BG:95}.
The three-loop term has been obtained, by a completely different method,
in~\cite{MR:00}.
Our calculation confirms this result.
It is interesting to note that the difference of the HQET
and QCD~\cite{LV:93,CR:00}
quark field anomalous dimensions $\tilde{\gamma}_Q-\gamma_q$
is gauge-invariant up to two loops~\cite{BG:91}
but not at three loops~\cite{MR:00}.

The renormalized heavy-quark propagator is 
\begin{eqnarray}
&&\left[\omega \tilde{S}_r(\omega; \mu = -2 \omega)\right]^{-1} =
1 + C_F \Sigma_F \frac{\alpha_s}{4\pi}
\nonumber\\
&&\quad{} +
C_F \left( C_F \Sigma_{FF} + C_A \Sigma_{FA} + T_f n_l \Sigma_{Fl} \right)
\left(\frac{\alpha_s}{4\pi}\right)^2
\nonumber\\
&&\quad{} +
C_F \left( C_F^2 \Sigma_{FFF} + C_F C_A \Sigma_{FFA} + C_A^2 \Sigma_{FAA} \right.
\nonumber\\
&&\qquad{} + \left.
C_F T_F n_l \Sigma_{FFl} + C_A T_F n_l \Sigma_{FAl} + (T_F^2 n_l)^2 \Sigma_{Fll}
\right) \left(\frac{\alpha_s}{4\pi}\right)^3
+ \cdots
\label{Sw}\\
&&\Sigma_{F} = - 4\,,\quad
\Sigma_{FF} = - 2 \zeta_2 a^2 + 12 \zeta_2 a - 18 \zeta_2 + 8\,,
\nonumber\\
&&\Sigma_{FA} = \left(\zeta_2+\frac{3}{8}\right) a^2
+ \left(-\zeta_2+3\right) a - 10 \zeta_2 - \frac{1537}{24}\,,\quad
\Sigma_{Fl} = 8 \zeta_2 + \frac{76}{3}\,,
\nonumber\\
&&\Sigma_{FFF} = - \frac{8}{3} \zeta_3 a^3
+ \left(24\zeta_3+8\zeta_2\right) a^2
+ \left(-72\zeta_3-48\zeta_2\right) a
+ 72 \zeta_3 + 72 \zeta_2 - \frac{32}{3}\,,
\nonumber\\
&&\Sigma_{FFA} = \left(4\zeta_3-\zeta_2\right) a^3
+ \left(-9\zeta_2-\frac{3}{2}\right) a^2
\nonumber\\
&&\quad{} + \left(-124\zeta_3+\frac{439}{3}\zeta_2-12\right) a
+ 264 \zeta_3 - 315 \zeta_2 + \frac{1537}{6}\,,
\nonumber\\
&&\Sigma_{FAA} = \left(-\frac{4}{3}\zeta_3+\zeta_2+\frac{29}{48}\right) a^3
+ \left(-\frac{1}{16}\zeta_4-\frac{19}{4}\zeta_3+\frac{15}{4}\zeta_2+\frac{197}{64}\right) a^2
\nonumber\\
&&\quad{} + \left(\frac{19}{3}\zeta_5+\frac{10}{3}\zeta_2\zeta_3
-\frac{175}{6}\zeta_4+\frac{130}{3}\zeta_3-\frac{187}{36}\zeta_2
+\frac{13921}{576}\right) a
\nonumber\\
&&\quad{} - 31 \zeta_5 + 22 \zeta_2 \zeta_3 + \frac{4603}{48} \zeta_{4}
- \frac{2057}{36} \zeta_3 - \frac{2167}{6} \zeta_2 - \frac{4176353}{3888}\,,
\nonumber\\
&&\Sigma_{FFl} = \left(32\zeta_3-\frac{128}{3}\zeta_2\right) a
- 24 \zeta_4 - 200 \zeta_3 + 120 \zeta_2 + \frac{179}{2}\,,
\nonumber\\
&&\Sigma_{FAl} = \left(-\frac{32}{3}\zeta_3+\frac{2}{9}\zeta_2-\frac{673}{72}\right) a
- 24 \zeta_4 + \frac{260}{9} \zeta_3 + \frac{2836}{9} \zeta_2 + \frac{184346}{243}\,,
\nonumber\\
&&\Sigma_{Fll} = \frac{160}{9} \zeta_3 - \frac{512}{9} \zeta_2 - \frac{31232}{243}\,.
\nonumber
\end{eqnarray}
For other normalization scales $\mu$, it can be obtained
by solving the renormalization group equation with~(\ref{gammaQ}).

Analytically continuing the coordinate-space propagator
from the half-axis $t>0$ to $t=-i\tau$, we have
\begin{eqnarray}
&&\log\left(
i \tilde{S}_r(\omega;\mu = \frac{2}{\tau}\,\mathrm{e}^{-\gamma_E})
\right) = C_F S^r_F \frac{\alpha_s}{4\pi}
\nonumber\\
&&{} + C_F \left( C_A S^r_{FA} + T_f n_l S^r_{Fl} \right)
\left(\frac{\alpha_s}{4\pi}\right)^2
\label{St}\\
&&{} + C_F \left( C_A^2 S^r_{FAA} + C_F T_F n_l S^r_{FFl}
+ C_A T_F n_l S^r_{FAl} + C_F (T_F n_l)^2 S^r_{Fll} \right)
\left(\frac{\alpha_s}{4\pi}\right)^3\,,
\nonumber\\
&&S^r_{F} = 4\,,\quad
S^r_{FA} = - \frac{3}{8} a^2
+ \left(4\zeta_2-3\right) a
- 12 \zeta_2 + \frac{1537}{24}\,,\quad
S^r_{Fl} = - \frac{76}{3}\,,
\nonumber\\
&&S^r_{FAA} = - \frac{29}{48} a^3
+ \left(\frac{1}{16}\zeta_4-\frac{5}{4}\zeta_3+\frac{5}{2}\zeta_2-\frac{197}{64}
\right) a^2
\nonumber\\
&&\quad{} + \left(-\frac{19}{3}\zeta_5-\frac{10}{3}\zeta_2\zeta_3
+\frac{175}{6}\zeta_4-64\zeta_3+\frac{304}{9}\zeta_2-\frac{13921}{576}\right) a
\nonumber\\
&&\quad{} + 31 \zeta_5 - 22 \zeta_2 \zeta_3 - \frac{4603}{48} \zeta_4
+ \frac{1089}{4} \zeta_3 - \frac{2533}{18} \zeta_{2} + \frac{4176353}{3888}\,,
\nonumber\\
&&S^r_{FFl} = 24 \zeta_4 + 104 \zeta_3 - \frac{1145}{6}\,,\quad
S^r_{Fll} = \frac{32}{3} \zeta_3 + \frac{31232}{243}\,,
\nonumber\\
&&S^r_{FAl} = \left(16\zeta_3-\frac{80}{9}\zeta_2+\frac{673}{72}\right) a
+ 24 \zeta_4 - \frac{556}{3} \zeta_3 + \frac{352}{9} \zeta_2 - \frac{184346}{243}\,.
\nonumber
\end{eqnarray}

\section{Heavy-light quark current}
\label{Current}

In order to calculate the anomalous dimension of the HQET heavy-light current,
we consider the bare proper vertex function of this current.
We put the light-quark momentum to zero.
The heavy quark carries residual energy $\omega<0$;
this is enough to ensure infrared convergence.
The integrals are of the same propagator type as discussed in Sect.~\ref{Dia}.
We follow the same automatic procedure as in Sect.~\ref{Prop}:
the diagrams are generated by QGRAF
(there are 237 of them),
processed by the Mathematica program,
traces and contractions are calculated with FORM,
and integrals -- with Grinder.
The exact $d$-dimensional result and its $\varepsilon$-expansion are available
at~\cite{Progdata}.

After re-expressing this bare vertex function via the renormalized quantities
$\alpha_s(\mu)$, $a(\mu)$, it becomes $\tilde{Z}_\Gamma$ times
the renormalized vertex, which is finite at $\varepsilon\to0$.
We have checked that the higher $1/\varepsilon$ poles in $\tilde{Z}_\Gamma$
are those dictated by the renormalization group.
The renormalization constant
$\tilde{Z}_j=\tilde{Z}_\Gamma\tilde{Z}_Q^{1/2}Z_q^{1/2}$
(with $\tilde{Z}_Q$ from Sect.~\ref{Prop} and $Z_q$ from~\cite{LV:93})
is gauge-invariant, as it should be for a gauge-invariant current $\tilde{\jmath}$.
The resulting  anomalous dimension reads
\begin{eqnarray}
&&\tilde{\gamma}_j = - 3 C_F \frac{\alpha_s}{4\pi}
\nonumber\\
&&{} + C_F \left[ C_F \left(-16\zeta_2+\frac{5}{2}\right)
+ C_A \left(4\zeta_2-\frac{49}{6}\right)
+ \frac{10}{3} T_F n_l \right]
\left(\frac{\alpha_s}{4\pi}\right)^2
\nonumber\\
&&{} + C_F \Biggl[
C_F^2 \left(-80\zeta_4-36\zeta_3+64\zeta_2-\frac{37}{2}\right)
\nonumber\\
&&\quad{} + C_F C_A
\left(-16\zeta_4+\frac{142}{3}\zeta_3-\frac{1184}{9}\zeta_2-\frac{655}{36}\right)
\nonumber\\
&&\quad{} + C_A^2
\left(-24\zeta_4-\frac{22}{3}\zeta_3+\frac{260}{9}\zeta_2+\frac{1451}{108}\right)
\nonumber\\
&&\quad{} + C_F T_F n_l
\left(-\frac{176}{3}\zeta_3+\frac{448}{9}\zeta_2+\frac{470}{9}\right)
\nonumber\\
&&\quad{} + C_A T_F n_l
\left(\frac{152}{3}\zeta_3-\frac{112}{9}\zeta_2-\frac{512}{27}\right)
 + \frac{140}{27} \left(T_F n_l\right)^2
\Biggr]
\left(\frac{\alpha_s}{4\pi}\right)^3 + \cdots
\label{gammaj}
\end{eqnarray}
This is the main result of this paper.
The two-loop term coincides with~\cite{JM:91,BG:91}.
The three-loop term with $(T_F n_l)^2$
is contained in the all-order large-$n_l$ result~\cite{BG:95}.
All the rest is new.

For the standard QCD  ($SU_c(3)$ colour group)
values  $C_F= 4/3, \, C_A = 3, \, T_F = 1/2\  $ 
eq.~(\ref{gammaj}) becomes
\begin{eqnarray}
\tilde{\gamma_{j}} =
&{-}& \frac{\alpha_s}{\pi}
{+}
\left[
-\frac{127}{72} 
+\frac{5}{36} n_l 
-\frac{7}{9}  \,\zeta_{2}
\right]\left(\frac{\alpha_s}{\pi}\right)^2
\nonumber\\
&{+}&
\quad
\left[
\frac{61}{192} 
+\frac{43}{324} n_l 
+\frac{35}{1296} n_l^2 
-\frac{343}{108}  \,\zeta_{2}
+\frac{49}{162} n_l  \,\zeta_{2}
\right. \nonumber \\ &{}& \left.
\phantom{+\left(\frac{\alpha_s}{\pi}\right)^3}
+\frac{89}{72}  \,\zeta_{3}
+\frac{83}{108} n_l  \,\zeta_{3}
-\frac{475}{54}  \,\zeta_{4}
\right]\left(\frac{\alpha_s}{\pi}\right)^3
\nonumber
\\
&=&
- \frac{\alpha_s}{\pi} +  (-3.04328 + 0.138889 \,n_l\,\,)
\left(\frac{\alpha_s}{\pi}\right)^2 
\nonumber
\\
&+& 
(-12.941 + 1.55406 \,n_l\,\, + 0.0270062 \,n_l^2\, )
\left(\frac{\alpha_s}{\pi}\right)^3
{}.
\label{jsut:QCD+N}
\end{eqnarray}

\section{Application: $f_B/f_D$}
\label{App}

As a simple application, let's consider the ratio $f_B/f_D$.
We have
\[
f_D = \frac{1}{\sqrt{m_c}}
\left[ 1 + c_1 \frac{\alpha_s^{(4)}(m_c)}{4\pi}
+ c_2^{(3)} \left(\frac{\alpha_s^{(4)}(m_c)}{4\pi}\right)^2
+ \cdots \right]
F^{(3)}(m_c)
+ \mathcal{O}\left(\frac{\Lambda_{\mathrm{QCD}}}{m_c}\right)\,,
\]
where~\cite{BG:95,G:98}
\begin{eqnarray}
&&c_1 = - 2 C_F\,,
\nonumber\\
&&c_2^{(n_l\,)}
= C_F \Biggl[ C_F \left(4\pi^2\log2-14\zeta_3-15\zeta_2+\frac{255}{16}\right)
\nonumber\\
&&\quad{} + C_A \left(-2\pi^2\log2+5\zeta_3+5\zeta_2-\frac{871}{48}\right)
\nonumber\\
&&\quad{} + T_F n_l \left(4\zeta_2+\frac{47}{12}\right)
+ T_F \left(-24\zeta_2+\frac{727}{18}\right)\Biggr]\,,
\label{Cg0}
\end{eqnarray}
and we have neglected corrections $\sim(m_c/m_b)^2$ in $c_2^{(n_l)}$
($m_{b,c}$ are the on-shell masses of $b$, $c$).
Similar formulas hold for $f_B$.
Running of the HQET matrix element $F^{(4)}(\mu)$ is governed
by the anomalous dimension~(\ref{gammaj}) with $n_l=4$;
$F^{(4)}(m_c)$ is related to $F^{(3)}(m_c)$ by the decoupling relation~\cite{G:98}
\begin{equation}
F^{(4)}(m_c) = \left[1 +
\frac{89}{36} C_F T_F \left(\frac{\alpha_s^{(4)}(m_c)}{4\pi}\right)^2
+ \cdots \right] F^{(3)}(m_c)\,.
\label{Dec}
\end{equation}

We obtain
\begin{eqnarray}
&&\frac{f_B}{f_D} = \sqrt{\frac{m_c}{m_b}} x^a
\Biggl[1 + r_1 (x-1) \frac{\alpha_s^{(4)}(m_b)}{4\pi}
\nonumber\\
&&{} + \left(r_{20} + r_{21} (x^2-1) + \frac{1}{2} r_1^2 (x-1)^2\right)
\left(\frac{\alpha_s^{(4)}(m_b)}{4\pi}\right)^2
\nonumber\\
&&{} + \mathcal{O}\left(\alpha_s^3,\frac{\Lambda_{\mathrm{QCD}}}{m_{c,b}}\right)
\Biggr]\,,
\label{fBD}
\end{eqnarray}
where
\begin{eqnarray*}
&&x = \frac{\alpha_s^{(4)}(m_c)}{\alpha_s^{(4)}(m_b)} \approx 1.56\,,\quad
a = - \frac{\tilde{\gamma}_{j0}}{2\beta_0} = \frac{6}{25}\,,\\
&&r_1 = - c_1
- \frac{\tilde{\gamma}_{j0}}{2\beta_0}
\left(\frac{\tilde{\gamma}_{j1}}{\tilde{\gamma}_{j0}} - \frac{\beta_1}{\beta_0}\right)
= \frac{56}{75} \zeta_2 + \frac{4403}{1875}\,,\\
&&r_{20} = \Delta c_2 + z_2 = \frac{8}{3} \zeta_2 + \frac{115}{27}\,,\\
&&r_{21} = - c_2^{(3)} + \frac{1}{2} c_1^2 + z_2
+ \frac{\tilde{\gamma}_{j0}}{4\beta_0} \left(
- \frac{\tilde{\gamma}_{j2}}{\tilde{\gamma}_{j0}}
+ \frac{\beta_2}{\beta_0}
+ \frac{\beta_1\tilde{\gamma}_{j1}}{\beta_0\tilde{\gamma}_{j0}}
- \frac{\beta_1^2}{\beta_0^2} \right)\\
&&\quad{} = \frac{8}{9} \pi^2 \log 2 + \frac{152}{9} \zeta_4 - \frac{254}{75} \zeta_3
+ \frac{272392}{16875} \zeta_2 + \frac{2071339}{281250}\,,
\end{eqnarray*}
$\beta$-function and the anomalous dimension are for 4 flavours,
$\Delta c_2=c_2^{(4)}-c_2^{(3)}=C_F T_F\left(4\zeta_2+\frac{47}{12}\right)$,
$z_2=\frac{89}{36} C_F T_F$ (see~(\ref{Cg0}), (\ref{Dec})).
Numerically,
\begin{eqnarray}
&&\frac{f_B}{f_D} = \sqrt{\frac{m_c}{m_b}} x^{6/25}
\Biggl[1 + 0.894 (x-1) \frac{\alpha_s^{(4)}(m_b)}{\pi}
\nonumber\\
&&{} + (3.788 x^2 - 0.799 x - 2.448) \left(\frac{\alpha_s^{(4)}(m_b)}{\pi}\right)^2
+ \mathcal{O}\left(\alpha_s^3,\frac{\Lambda_{\mathrm{QCD}}}{m_{c,b}}\right)
\Biggr]\,.
\label{fBDnum}
\end{eqnarray}
The coefficient of $(\alpha_s(m_b)/\pi)^2$ is about 5.5.
Note that the power corrections $\sim(m_c/m_b)^n$ are omitted
from the NNL term, so that this formula does not apply at $m_c\sim m_b$.
Unfortunately, lattice and sum-rules estimates show that the $\Lambda_{\mathrm{QCD}}/m_c$
correction to $f_D$ is large.

\section{Conclusion}
\label{Conc}

We have calculated the HQET heavy-quark self-energy up to three loops,
in an arbitrary covariant gauge.
The exact $d$-dimensional result and its $\varepsilon$-expansion
are available at~\cite{Progdata}.
Upon renormalizing the vertex function, 
we have obtained the heavy-quark field anomalous dimension~(\ref{gammaQ}),
in agreement with~\cite{MR:00},
and the renormalized heavy-quark propagator~(\ref{Sw}), (\ref{St}).
We have calculated the vertex function of the HQET heavy-light current
with zero momentum of the light quark, up to three loops,
in an arbitrary covariant gauge.
The results are available at the same site.
From it, we have obtained the anomalous dimension
of the heavy-light current~(\ref{gammaj}).
This is the main result of this paper.
We have derived the next-to-next-to-leading perturbative correction
to $f_D/f_B$~(\ref{fBD}).

We are grateful to R.~Sommer for a discussion of lattice calculations
of $f_B$ and careful reading of the text.
The work of KCh was supported by the DFG
Sonderforschungsbereich SFB/TR9 ``Computational Particle Physics'', by
INTAS (grant 00-00313) as well as by the European Union under contract
HPRN-CT-2000-00149.  AG acknowledges support by the German Ministry
for Research BMBF, Contract No.\ 05HT1VKB1.

\end{document}